# Modulator-Based, High Bandwidth Optical Links for HEP Experiments

D. G. Underwood, G. Drake, W. S. Fernando, R. W. Stanek

*Abstract*— **As a concern with the reliability, bandwidth and mass of future optical links in LHC experiments, we are investigating CW lasers and light modulators as an alternative to VCSELs. These links will be particularly useful if they utilize light modulators which are very small, low power, high bandwidth, and are very radiation hard. We have constructed a test system with 3 such links, each operating at 10 Gb/s. We present the quality of these links (jitter, rise and fall time, BER) and eye mask margins (10GbE) for 3 different types of modulators: LiNbO3-based, InP-based, and Si-based. We present the results of radiation hardness measurements with up to ~$10^{12}$ protons/cm$^2$ and ~65 krad total ionizing dose (TID), confirming no single event effects (SEE) at 10 Gb/s with either of the 3 types of modulators. These optical links will be an integral part of intelligent tracking systems at various scales from coupled sensors through intra-module and off detector communication. We have used a Si-based photonic transceiver to build a complete 40 Gb/s bi-directional link (10 Gb/s in each of four fibers) for a 100m run and have characterized it to compare with standard VCSEL-based optical links. Some future developments of optical modulator-based high bandwidth optical readout systems, and applications based on both fiber and free space data links, such as local triggering and data readout and trigger-clock distribution, are also discussed.**

***Index Terms*—Optical modulators, Radiation effects, High Energy Physics Instrumentation, Vertical Cavity Surface Emitting Lasers**

## I. Introduction

HEP experiments need a reliable, high bandwidth, low power technology for data transmission. [1] [2][3] There is usually no access to repair any failed devices inside the experimental detector for a year. Many of the current laser-based optical links at the LHC failed for many reasons (known and unknown). The future of optical links is light modulators, and we are investigating how these can be integrated into HEP experiments.

David Underwood is with High Energy Physics, Argonne National Laboratory, 9700 S. Cass Ave, Argonne, Il, 60439 USA, (e-mail: dgu@hep.anl.gov)

Gary Drake is with High Energy Physics, Argonne National Laboratory, 9700 S. Cass Ave, Argonne, Il, 60439 USA

W. Fernando is with High Energy Physics, Argonne National Laboratory, 9700 S. Cass Ave, Argonne, Il, 60439 USA

R. Stanek is with High Energy Physics, Argonne National Laboratory, 9700 S. Cass Ave, Argonne, Il, 60439 USA

Some advantages of modulators over the current Vertical Cavity Surface Emitting Laser (VCSEL) based links are:

- Low Bit Error Rates: simpler error correction code would be required.
- Improved radiation hardness: radiation damage generates defects in the p-n junction of the VCSEL and increases the laser diode threshold and decreases its optical efficiency. Hence we expect modulators to outperform VCSELs in reliability and radiation hardness.
- Higher data rates: VCSEL speeds saturate at 10 Gb/s. The optical modulators would easily work at 10 Gb/s and commercially available ones operate beyond 40 Gb/s.
- Low power consumption: modulators are driven with voltage at a low current and consume less current than directly modulated laser transmitters.

CW lasers also have many advantages in an HEP readout system. They have a simpler junction structure than VCSELs, with lower current densities, which improves reliability. They can be used near the data modulators in moderate radiation environments, or moved off-detector with the light carried by fiber, for higher radiation applications. CW laser light can be split to multiple light modulators, reducing power overhead. The turn on-turn off issues with VCSELs at GHz rates are eliminated, and control of modulator turn-on turn off is much more straightforward. [4][5]

## II. Modulators Under Study

Our group has worked with Lithium Niobate modulators at 1 to 10 Gb/s for some time. [6][7][8][9] There are versions of these that work up to 40 Gb/s on a single fiber, and they have been tested to be radiation resistant to the levels required by LHC experiments. However, they are not suitable for many applications due to the large physical size, the mass, and the high drive voltage required, typically of ~5 V into 50 ohms. We have been interested in integrated silicon optics and have done radiation tests of Ge doped Si modulator base material to levels of $10^{17}$ electrons. [10]

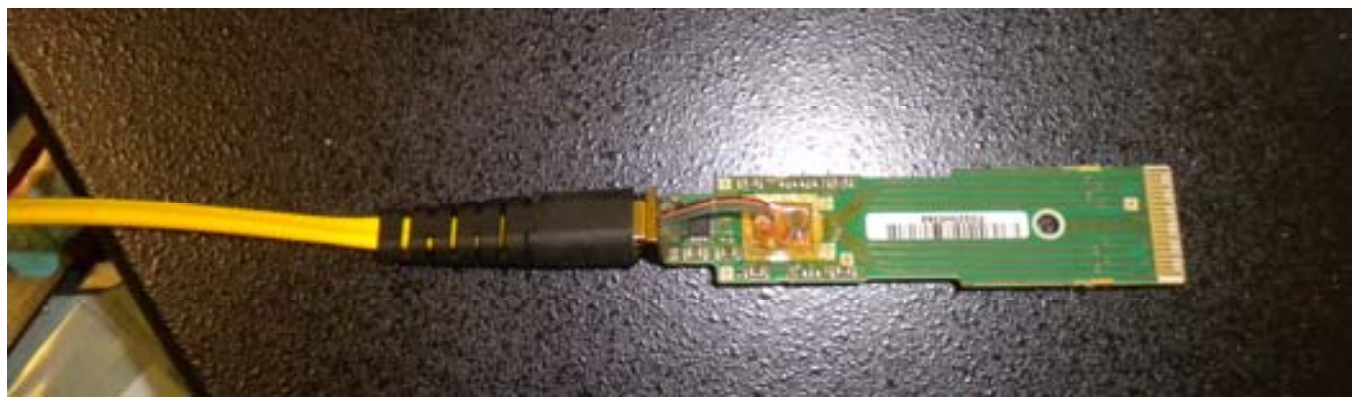

Fig.1. Example of Molex / Luxtera unit as a QSFP plugin with shell removed for visibility.

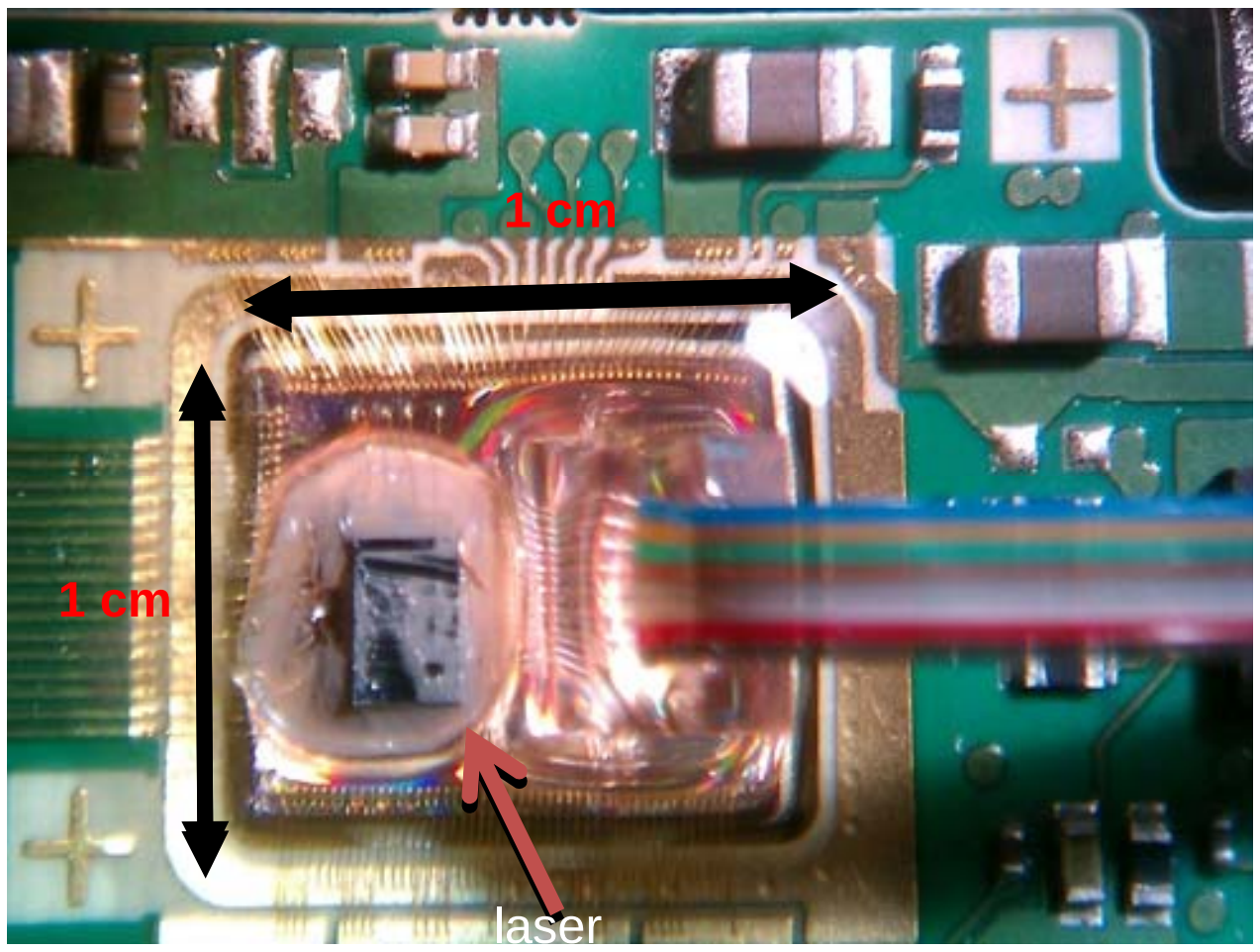

Fig. 2. A Si MZI based 4x10Gb/s from Luxtera/Molex fully integrated transceiver in silicon. This integrated optical chip has 4 transmitters of 10 Gb/s, and 4 receivers.

More recently we have focused on two other paths, commercial silicon integrated optics devices which might be usable either directly or with small modifications, an example of which is shown in Figs.1 and 2, and InP modulators which are expected to be more radiation-resistant than silicon, due to the band structure of the material, and which are also available commercially and work up to 40 Gb/s.[11] Example devices are shown in Figs. 3 and 4.

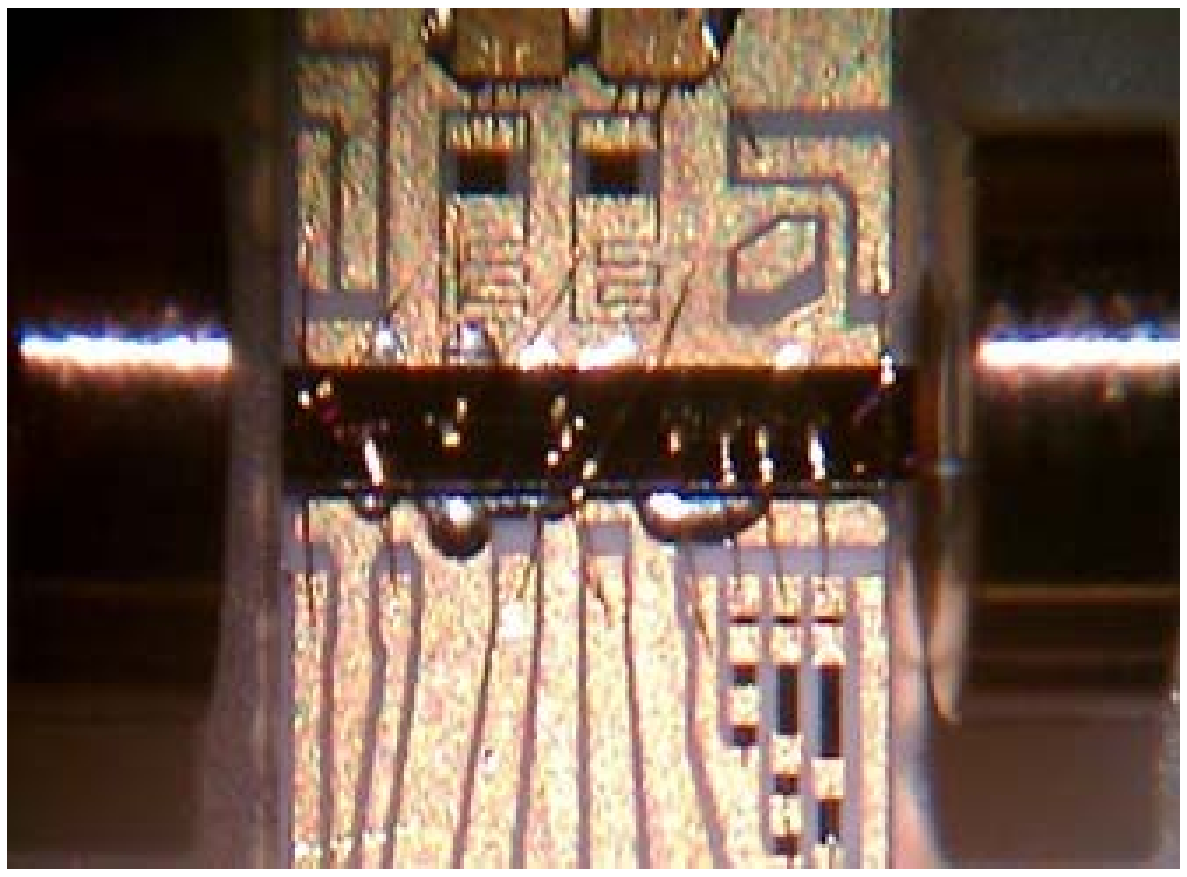

Fig 3. InP-based 10Gb/s device. This is a close-up of the actual modulator part (horizontal dark band. The actual modulator is quite small, roughly 3 mm long.

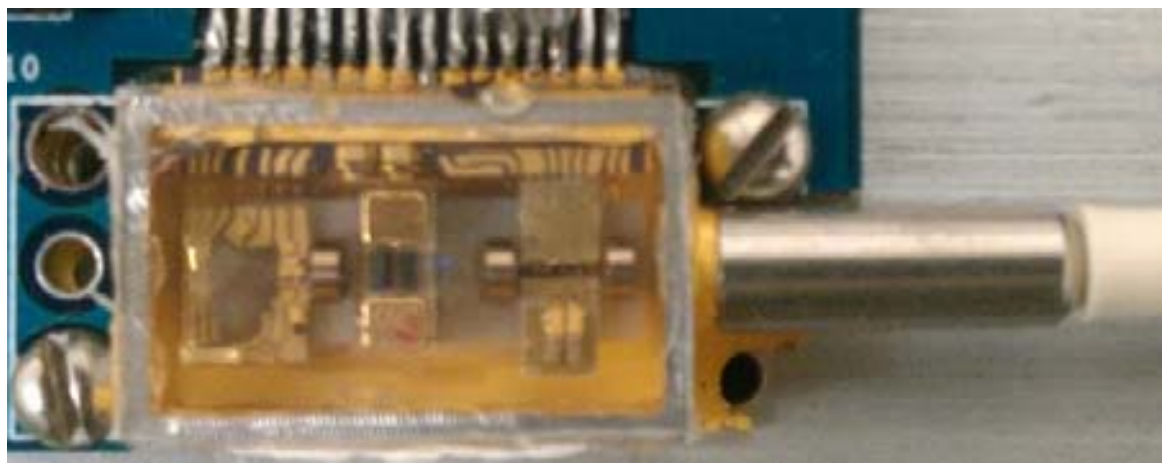

Fig.4. InP Mach-Zhender modulator. This small package has laser, lenses, polarizers, programmable attenuator, MZ modulator,

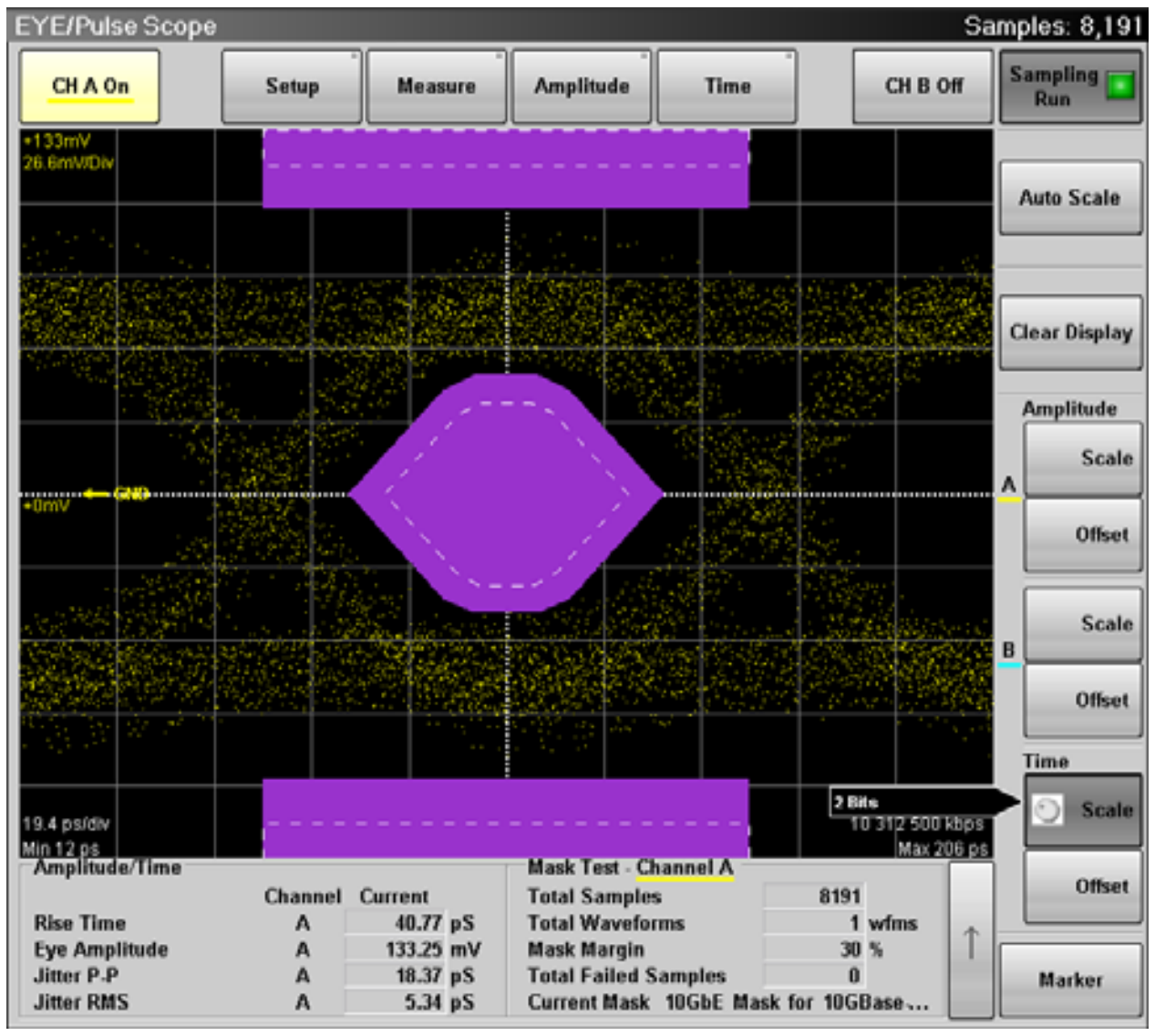

Fig. 5. Eye diagram of a LiNbO3 modulator at 10 Gb/s with an Anritsu communications analyzer showing IEEE802.3ae limits and data.

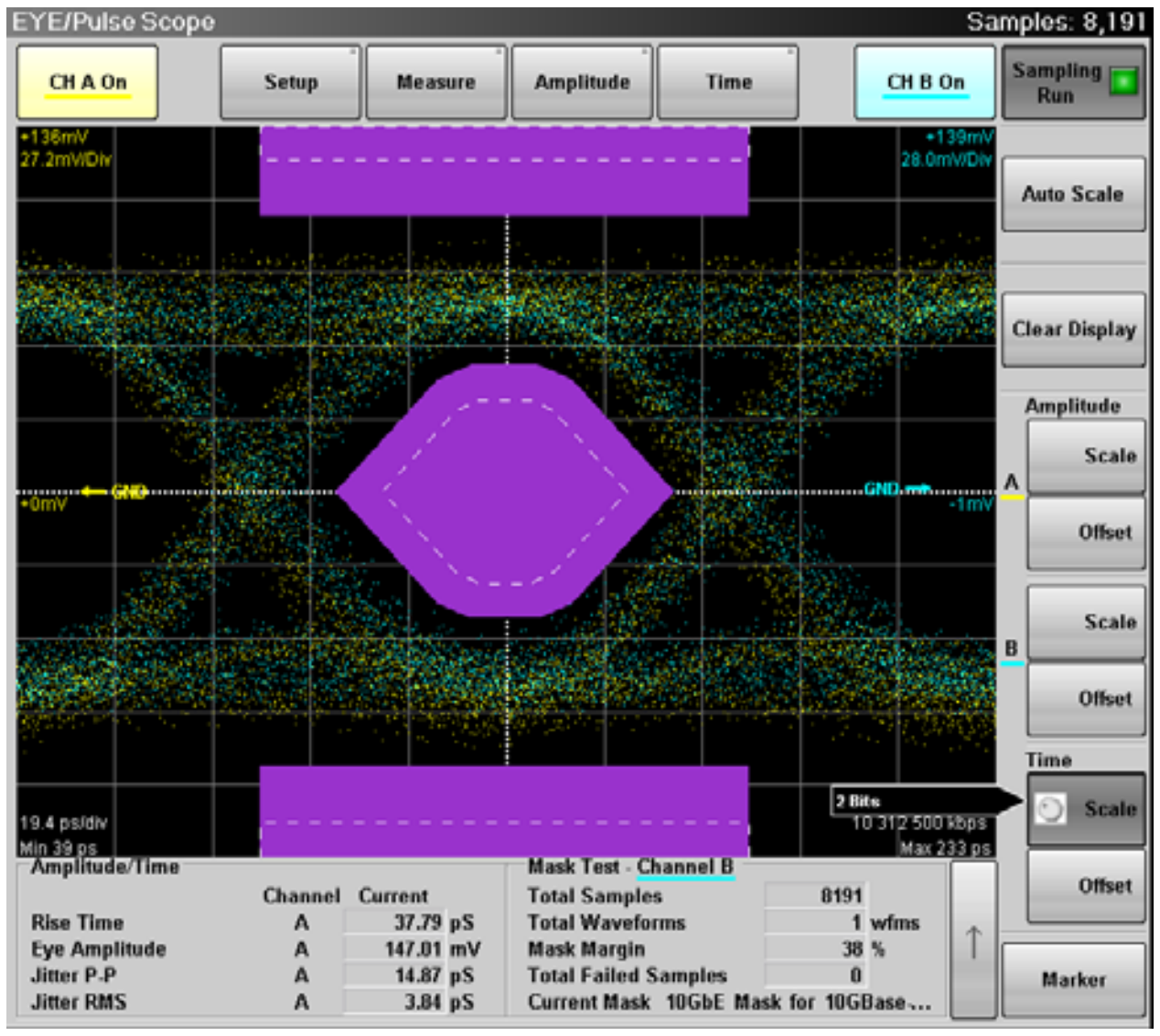

Fig. 6. Eye diagram of an InP modulator at 10 Gb/s with an Anritsu communications analyzer showing IEEE802.3ae limits and data.

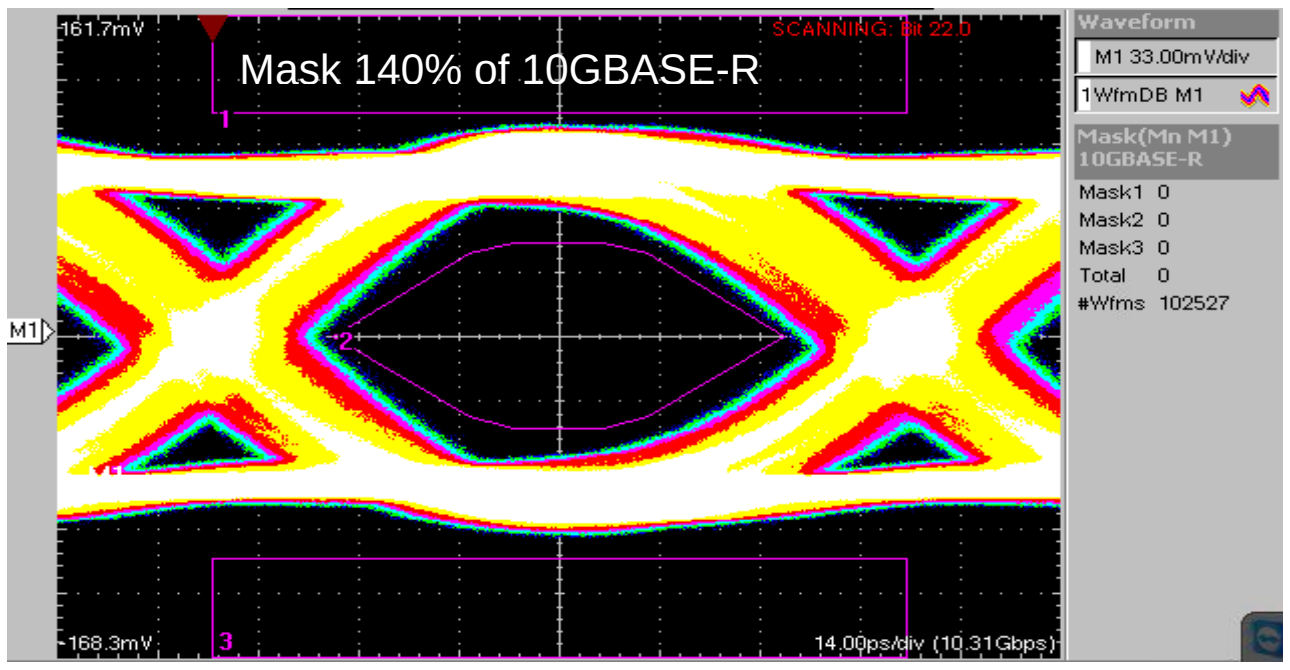

Fig. 7.  Eye diagram of the Luxtera/Molex modulator at 10 Gb/s.

## III. Bench Tests of Modulators

We did bench tests of all 3 modulator technologies at 10 Gb/s. All satisfied IEEE802.3ae requirements on the eye diagram as shown in Figs. 5-7. The BER was found to be quite sensitive to the quality of the SMA cables used. We did a bench test of the Molex / Luxtera QSFP with 200 m of fiber, as would be used in ATLAS. Bathtub curves run up to $10^{-15}$ predict the eye opening *vs* bit error rate based on the jitter & noise. The predicted eye opening from ANL tests is a BER of $10^{-18}$ as seen in Fig.8. Molex and Luxtera have done actual tests to $10^{-18}$ with no errors, which extrapolate to even better BER.

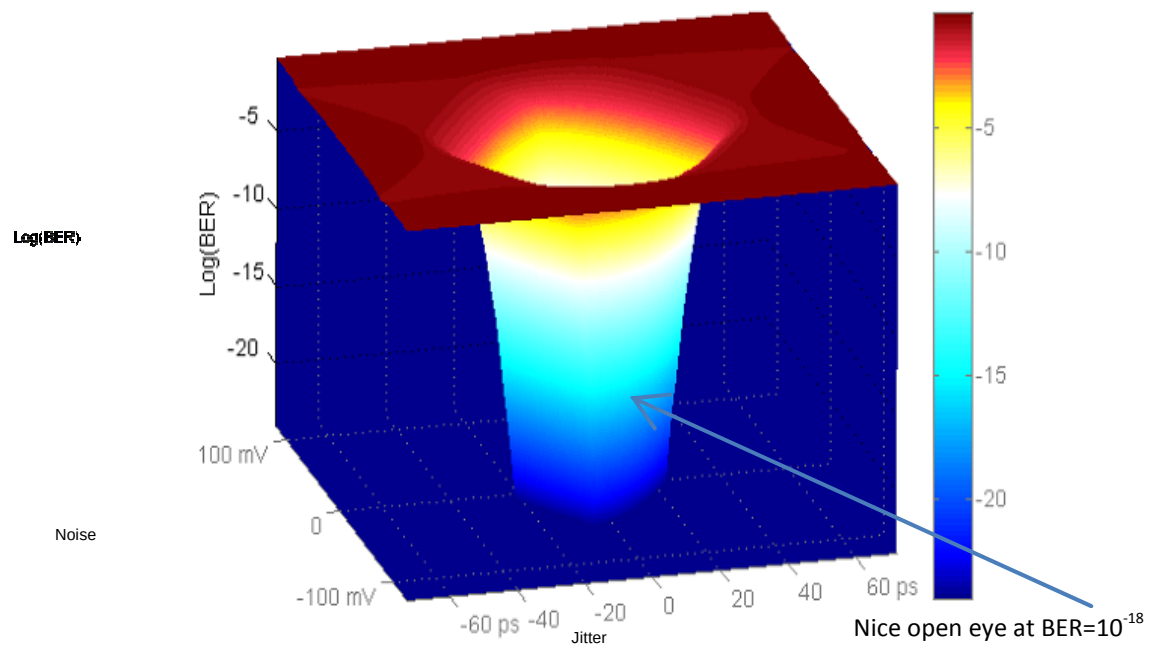

Fig. 8. Bathtub plot of BER test done at ANL to $10^{-15}$ showing extrapolation to $10^{-18}$. The clearance around the standard eye diagram limit which contains limits on amplitude variation and jitter is not visible in this view. Molex / Luxtera did an actual test to $10^{-18}$.

For a link running at 10 Gb/s an error rate of $10^{-18}$ is 1 error in 1000 days. While a BER of $10^{-12}$ implies 900 errors per day. The current VCSEL link specification for many experiments is $10^{-12}$.

## IV. Radiation Tests of Modulators

We tested 3 different types of modulators at the cyclotron at Massachusetts General Hospital (proton beam) with a varying energy from 207 to 85 MeV. Fig. 9 shows a schematic of the layout and interconnection and Fig. 10 shows the setup in the beamline. The energy deposit was well above that from minimum ionizing (MIP).

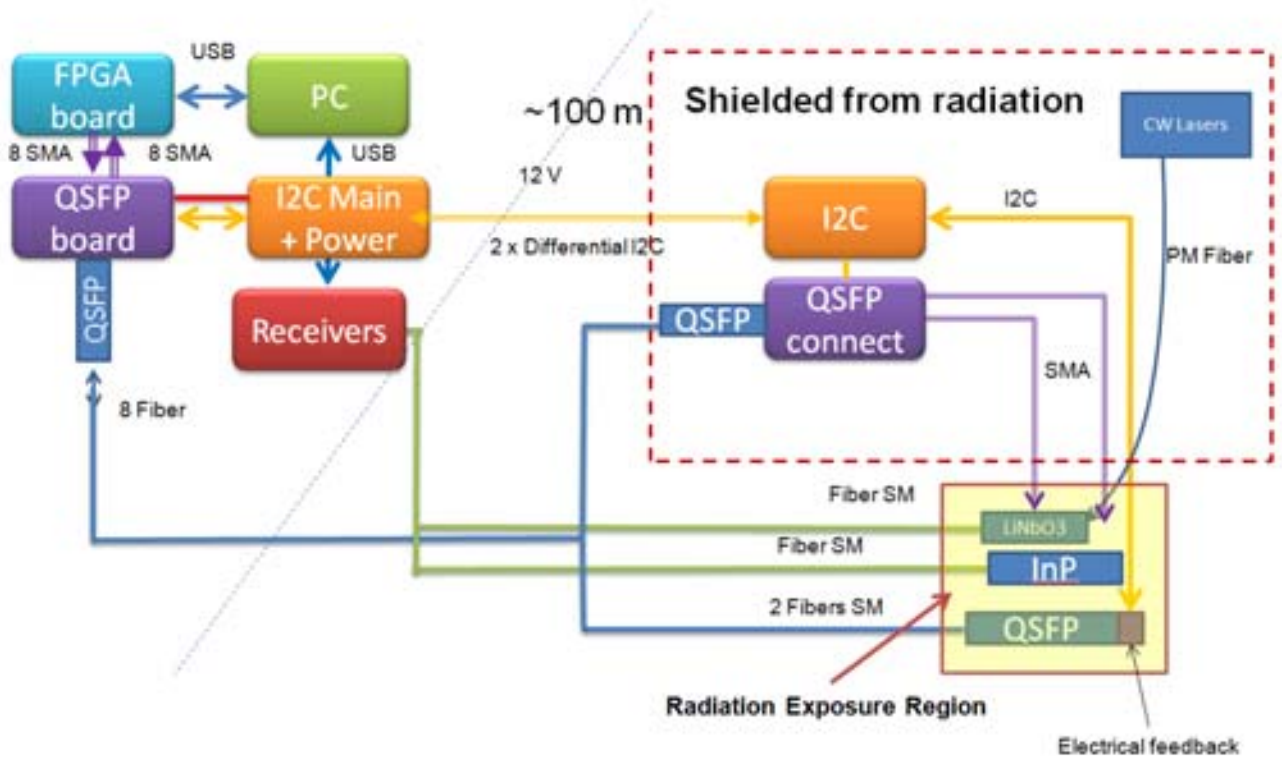

Fig. 9. Electronic / optical link setup used during the proton exposure of the Molex QSFP module at Mass. Gen. Hospital.

We tested a silicon integrated optics unit made by Molex / Luxtera, shown in Figs 1 and 2, and a commercial InP modulator with cables to an external driver, device shown in Figs.3 and 4, and a standard LiNO3 modulator used by telecom companies. All 3 were run with pseudo random data at 10 Gb/s continuously during the radiation exposure. The Molex / Luxtera QSFP were run with 200 m fiber, as would be used in an ATLAS application.

The total ionizing dose, TID, was 64 krad, from a fluence of 8 x $10^{11}$ p/cm$^2$. There were no single event upsets (SEU). The bit error rate (BER) during the exposure was < 8.9 x $10^{-15}$ measured during the experiment which was time limited.

The requirements for ATLAS-LHC TileCal would be 45 krad, NIEL 1.5 x $10^{12}$ 1 MeV neutron/cm$^2$, and with a larger number of devices tested.

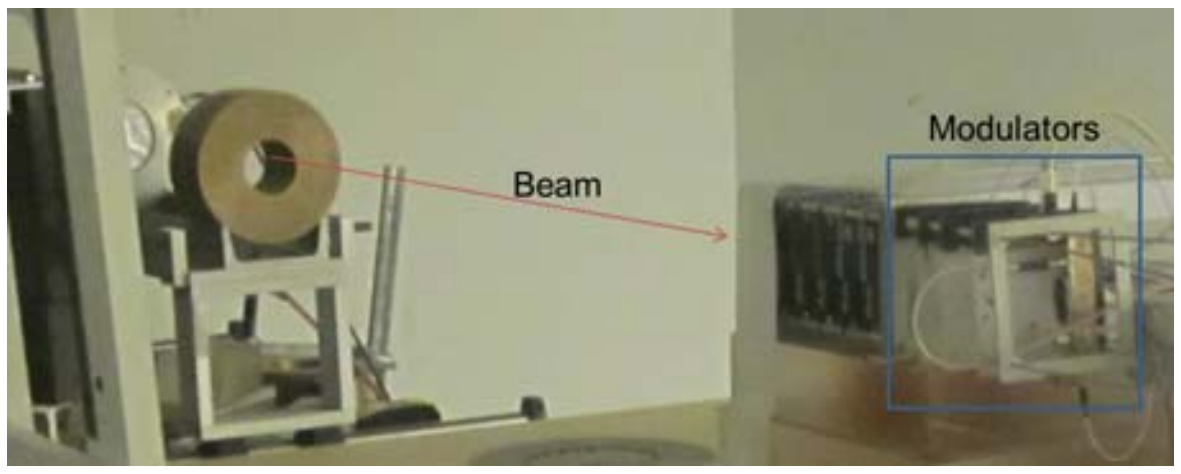

Fig. 10.  Test setup in the beam at Massachusetts General Hospital.

While neither the silicon integrated optical chip nor the InP modulator showed any problems due to irradiation, the microcontroller on the printed circuit board of the commercial QSFP did not survive. It is used to initiate the operation of the optical system, and it could not be restarted after a power down. The microcontroller uses flash memory. There are papers in the literature showing flash memory failures at roughly 20 krad, and a PIC (Microchip) microcontroller failure at around 20 krad. [12] Modern FPGAs work routinely at over 1000 times the microcontroller radiation levels. [13]

The InP MZ modulator also survived the irradiation with no SEE and no obvious damage.

We did two irradiations of Molex / Luxtera devices at the 3 MeV Argonne electron van de Graaff as seen in Fig.11. These were to at least 100 krad each. One channel of transmission and one channel of the receiver were in operation during the entire exposure. We used water cooling, and a cooled airflow. We also used a 6 mm thick brass shield in an attempt to shield the microcontroller while exposing the optical chip. As in the case of the proton exposure, the microcontroller appears to have been damaged by the radiation and would not restart after power down and, as expected with electrons, no SEE were seen.

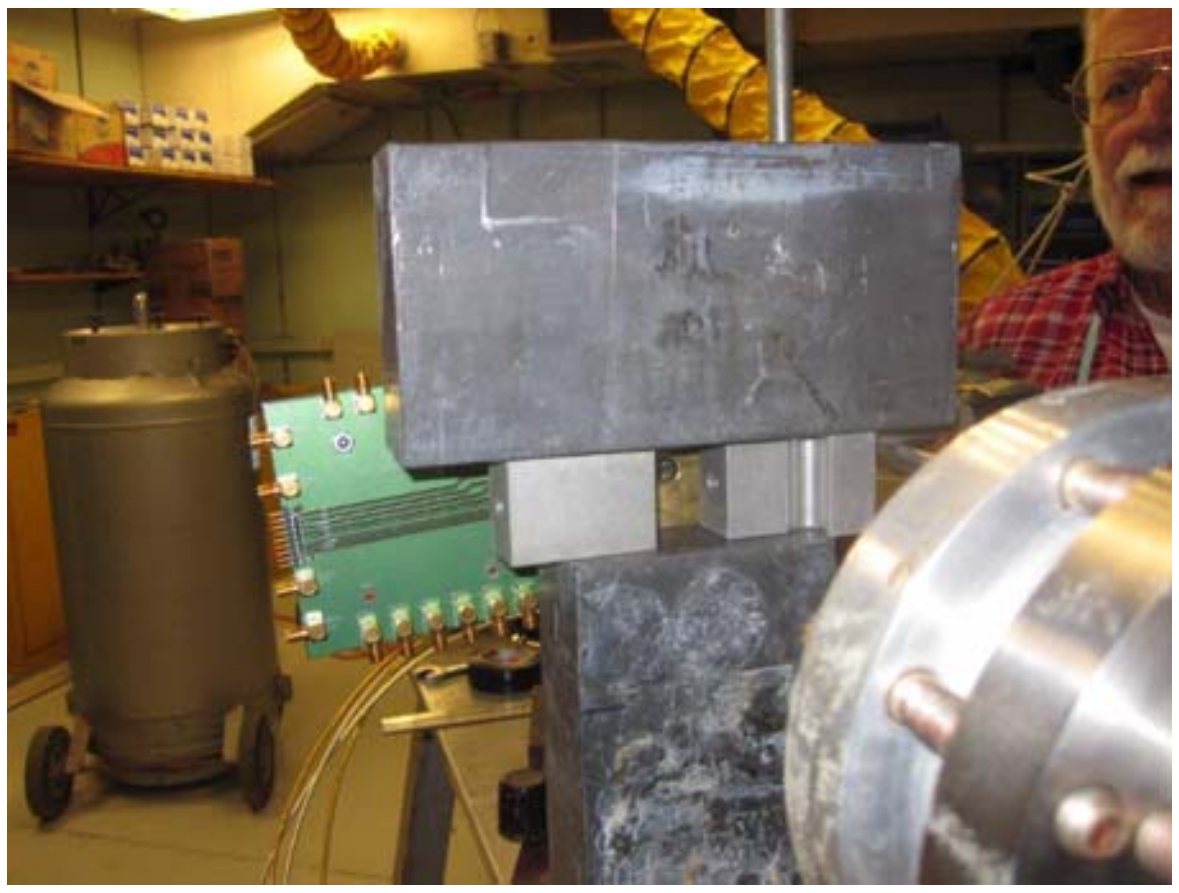

Fig. 11. Test setup at ANL 3 MeV van de Graaff. The modulator was in continuous operation during the 100 krad runs.

## V. Other Uses of Modulators

For on-board triggering between tracking layers, technology such as 3D silicon may be useful up to of order 1 mm, and conventional readout may be useful on relatively large scales, but there is a window of opportunity for layers separated by of order cm. Small light modulators with small prisms would allow communication between layers, or between 3D double layers (Fig.12).

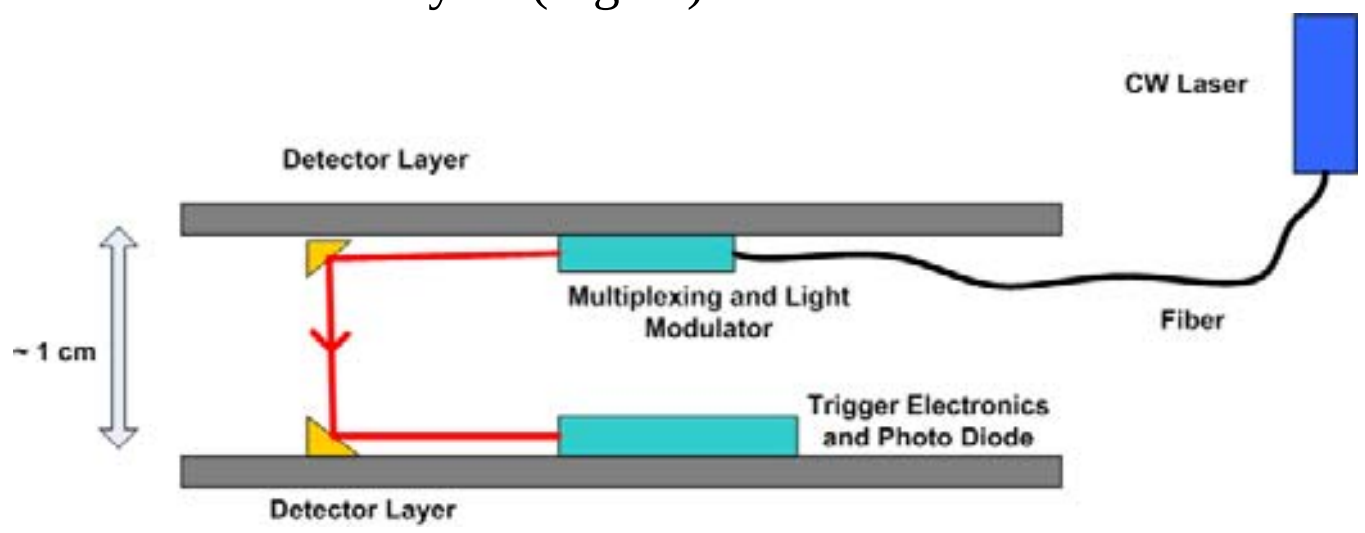

Fig. 12. Data path for on-board tracking trigger which could couple 2 planes of 3D doublets.

## VI. Conclusions

We are in the process of evaluating various forms of optical modulators for use in HEP experiments. The future of optical links in general is optical modulators, and it is clear from the literature and from our preliminary studies of radiation hardness and bit error rate that this is the most promising path for HEP data links.

Commercial Integrated Optics Chips are a promising form of Modulators. Some of the features:

- Speed
  - 10 Gb/fiber commercial integrated optics
  - 40 Gb/fiber with some commercial units
- Laser reliability
  - Either CW laser onboard
  - Or displace laser outside detector
  - Different junction structure than VCSELs
- Low Bit Error Rate
  - $10^{-18}$ *vs* typical $10^{-12}$ for current systems
  - Simplified error correction schemes
- Low power-
  - One CW laser split many ways
  - Modulators are very efficient
  - Short electrical paths – no cable drivers
  - Low voltage drivers – not current
- Rad hard optical parts-
  - We have thoroughly tested silicon integrated optics for 64 krad application
  - Modulator parts should work at much higher levels
  - Optical part is expected to work at multi-Mrad levels.

## Appendix

Appendixes, if needed, appear before the acknowledgment.

## Acknowledgment

We thank the HEP electronics staff for assistance in building and modifying circuitry. We thank Ethan Cascio at MGH for his effort in providing a beam for us, and we thank Kevin Quigley and Sergei Chemerisov at the ANL van de Graaff facility for providing their services.